\documentclass[aip,apl,twocolumn,showpacs,superscriptaddress,groupedaddress]{revtex4}
\usepackage[pdftex]{graphicx}
\usepackage[cmex10]{amsmath}
\usepackage{amssymb}
\begin{document}

\title{Relaxation-free and inertial switching in synthetic antiferromagnets subject to super-resonant excitation}

\author{B. C. Koop}
\affiliation{Royal Institute of Technology, 10691 Stockholm, Sweden}
\author{T. Descamps}
\affiliation{Royal Institute of Technology, 10691 Stockholm, Sweden}
\author{V. Korenivski}
\affiliation{Royal Institute of Technology, 10691 Stockholm, Sweden}
\date{\today}

\markboth{Journal of \LaTeX\ Class Files,~Vol.~13, No.~9, September~2014}%
{Shell \MakeLowercase{\textit{et al.}}: Bare Demo of IEEEtran.cls for Journals}

\begin{abstract}
Applications of magnetic memory devices greatly benefit from ultra-fast, low-power switching. Here we propose how this can be achieved efficiently in a nano-sized synthetic antiferromagnet by using perpendicular-to-the-plane picosecond-range magnetic field pulses. Our detailed micromagnetic simulations, supported by analytical results, yield the parameter space where inertial switching and relaxation-free switching can be achieved in the system. We furthermore discuss the advantages of dynamic switching in synthetic antiferromagnets and, specifically, their relatively low-power switching as compared to that in single ferromagnetic particles. Finally, we show how excitation of spin-waves in the system can be used to significantly reduce the post-switching spin oscillations for practical device geometries. 

\end{abstract}


\maketitle

Magnetic information storage has been the topic of extensive research for many decades. Memory devices are based on magnetization switching between two ground states of a nanomagnet and the ability via readout to distinguish these ground states as digital memory states "0" and "1". The most recent and most promising implementation of magnetic storage is magnetic random access memory (MRAM), the performance of which depends on optimizing its speed and power consumption. The speed of a memory device, desired to be as high as possible, has three main components: access-time, write-time, and read-time. MRAM, by design, has fast access and readout times, on par with the fastest semiconductor memories. It is optimization of the write time, desired to be as short as possible, and the write power, desired to be as low as possible, that can enhance the MRAM technology. This paper details a method of achieving these objectives using dynamic switching of synthetic antiferromagnetic (SAF) particles subject to ultrafast field pulses.   \par

SAF particles are widely used as storage elements in MRAM, traditionally with quasi-static (10 ns range) magnetic switching known as toggling \cite{TMRAM3,TMRAM4}. Dynamic, sub-ns range magnetic switching in the system has been investigated using several approaches: particle-shape optimization \cite{shape1,shape2}, write field- or write current-shaping \cite{H-shape1,I-shape1}, creating magnetic asymmetries \cite{asym1,asym2,asym3}, thermal activation \cite{thermal1,thermal2,combi1}, and non-uniform ground states \cite{nonuniform1,nonuniform2}. Most of the mentioned work explored combinations of these physical parameters. It has recently been predicted that ultrafast switching of single layers \cite{zpulse_SL}, antiferromagnets \cite{zpulse_AF} and SAFs \cite{Theory} can be achieved through the application of field pulses applied along the magnetic hard axes. In this paper we extend and detail our recent analytical prediction of ultrafast relaxation-free and inertial switching in SAF subject to an ultrafast field pulse applied normal to the SAF plane \cite{Theory}. The analytical results, obtained within certain realistic approximations, are verified using a detailed micromagnetic analysis free from such approximations. Additionally, we show that spin-wave mediated damping, found to be significant for a specific range of excitation parameters, can be used to further improve the speed and reliability of SAF switching.

%

\begin{figure}
\includegraphics[width=0.9\linewidth]{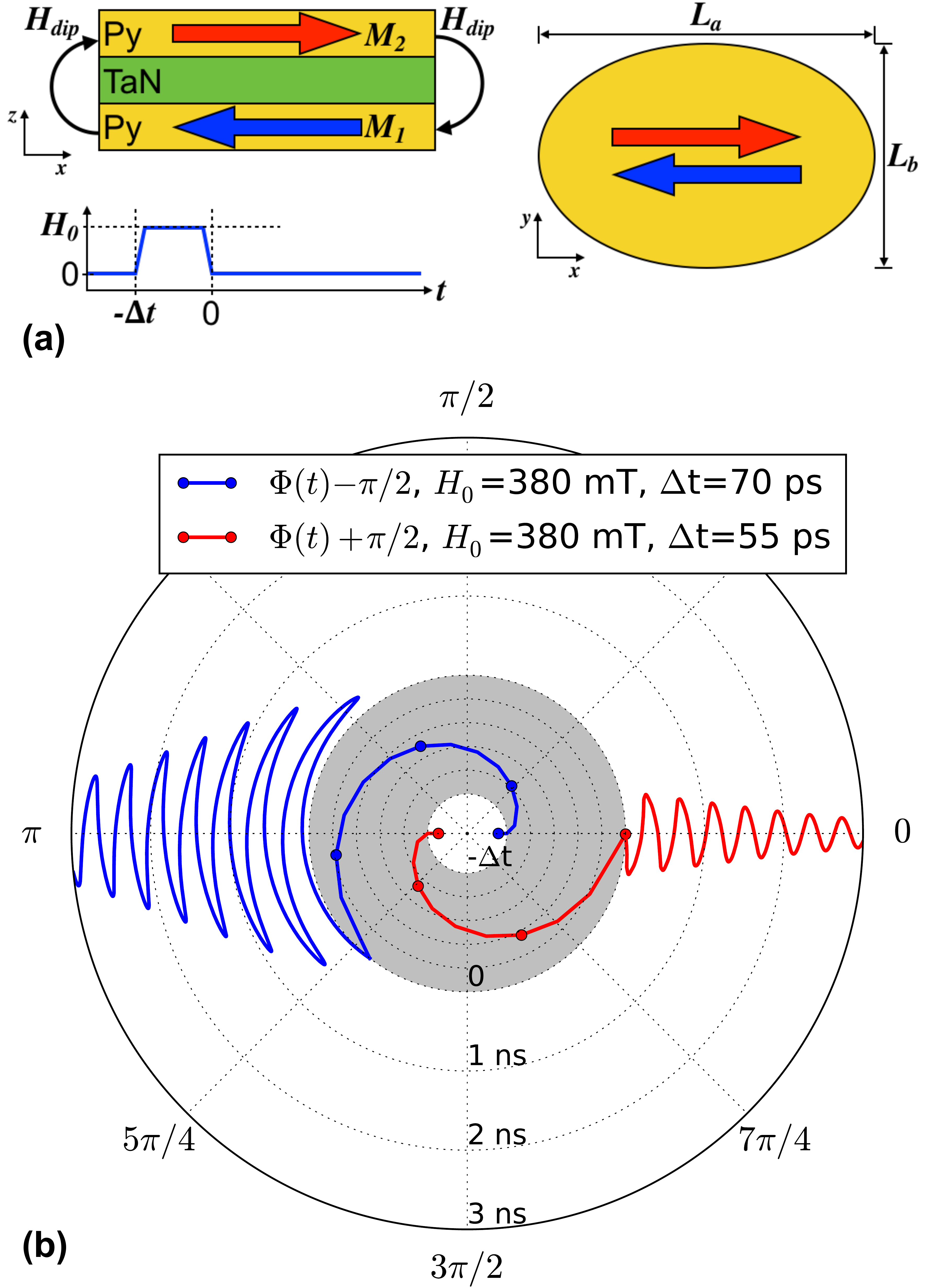}
\caption{\label{fig:fig1} (a) A schematic of the device configuration seen along the $y$-axis (top-left) and $z$-axis (right). Bottom left panel shows the notations used for the excitation field. (b) Time evolution of the magnetization excited by a 380 mT field pulse of two different durations. The radial scale represents time. The gray area denotes the magnetization evolution during the action of the pulse, over $t=[-\Delta t,0]$ range for each curve. Note that, due to the different pulse duration for the two traces, the respective timescale of the gray area is different. The amount of free relaxation after the pulse is switched off depends on the magnetization angle at that point in time, different for different pulse durations. }
\end{figure}

The SAF system considered in this paper consists of two closely separated thin elliptical magnetic disks. The interlayer exchange coupling is taken to be zero, which is appropriate for TaN-spacers in typical experimental SAF's \cite{TaN_spacer}. The two magnetic layers are dipole-coupled through the edge stray fields, which results in an antiparallel alignment of the layers' magnetization, as shown in figure \ref{fig:fig1}a. Due to the elliptical lateral shape and the small thickness, there are two ground states, with the magnetization aligned along the easy axis (EA) of the particle: $[\varphi_1,\varphi_2]  = [0,\pi]$ and $[\varphi_1,\varphi_2]  = [\pi,0]$, where $\varphi_i$ is the angle of the average layer's magnetization $M_i$ with respect to the $x$-axis. As shown in \cite{Theory} the criteria for relaxation-free switching under square-pulse excitation perpendicular to the layers are that the pulse amplitude $H_0$ and duration $\Delta t$ fulfill
\begin{eqnarray}
H_0/2\pi M_s &\gg& \sqrt{N_y-N_x-(\gamma_y-\gamma_x)},\nonumber\\
\label{eq:pulse_crit}\\
\Delta t \omega_0 &\ll& \pi/\sqrt{N_y-N_x-(\gamma_y-\gamma_x)},\nonumber
\end{eqnarray}
which leads to the following parametric conditions:
\begin{eqnarray}
\frac{\Delta m_z}{2\sqrt{N_y-N_x-(\gamma_y-\gamma_x)}} &\ll& \Delta\Phi \approx \frac{H_0\Delta t \omega_0}{4\pi M_s}\nonumber\\
\nonumber\\
\frac{1}{\omega_0}\left.\frac{d\Phi}{dt}\right|_{t=\Delta t} &=& -\frac{1}{2}m_z\nonumber.
\end{eqnarray}
Here $\omega_0=8\pi\mu_BM_s/\hbar$ and $\Phi=(\varphi_1+\varphi_2)/2$; $N_i$, $\gamma_i$ are the in-plane demagnetizing and dipole-coupling coefficients, respectively; $\Delta$ denotes the difference in the respective quantity ($Q$) at the start and the end of the pulse ($\Delta Q = Q(t=t_{start})-Q(t=t_{end})$).
This means that when the criteria of (\ref{eq:pulse_crit}) are satisfied, both the out-of-plane components of the magnetization as well as the in-plane angular velocity of the magnetization at the end of the excitation are minimal.
Therefore, switching under such specific excitation, with (\ref{eq:pulse_crit}) satisfied and $\Delta\Phi\approx\pi$, leads to \textit{no oscillations} in the final switched-to state of the SAF particle.
This can significantly reduce the effective switching time as well as enhance the stability of the destination ground state.

We focus on samples with lateral dimensions varied between 24 and 120 nm and layer thicknesses of 5 nm. The lateral aspect-ratios of the ellipses ($L_a/L_b$) range from 1.1 to 1.3. For these samples the minimum pulse-strength (using the criteria of eqs. \ref{eq:pulse_crit}) should be expected to range from 10 to 100 mT for pulse-durations of 1 ns down to 100 ps, respectively. Since the dynamic switching aimed at is by short, 100 ps range pulses, the relatively high field-amplitudes needed still result in modest (order of 1 nJ) effective pulse-energies for typical device geometries.

We next use fine-meshed micromagnetic simulations to verify the analytical predictions and test their limits. For this we use the OOMMF package \cite{oommf}. In the simulations the magnetization is first relaxed to one of the antiparallel ground states of the system. From this ground state a field pulse is applied normal to the film-plane with given strength ($H_0$) and duration ($\Delta t$) and the rise- and fall-times of 5 ps. The excitation layout is shown in figure \ref{fig:fig1}a. The thicknesses of the layers were kept constant (5 nm) and the lateral dimensions were varied. We have used a 1x1x5 nm$^3$ mesh throughout. 5 nm meshing in the z-direction was to reduce the simulation time. We have verified that a finer mesh in \textit{z}, down to 1 nm, does not influence the results.

The typical time evolution of the magnetization is shown in figure \ref{fig:fig1}b for a 33x30 nm SAF particle. For fast pulses (dynamic excitation) the profile of the magnetization relaxation strongly depends on the angle $\Phi$ it forms with respect to the easy axis at the end of the field pulse ($t=0$ s), while the rotational velocity $d\Phi/dt$ during the pulse is determined by the pulse strength $H_0$. The relaxation pattern therefore depends sensitively on the pulse magnitude and duration, in addition to the resonant properties of the SAF particle itself.

\begin{figure}
\includegraphics[width=0.9\linewidth]{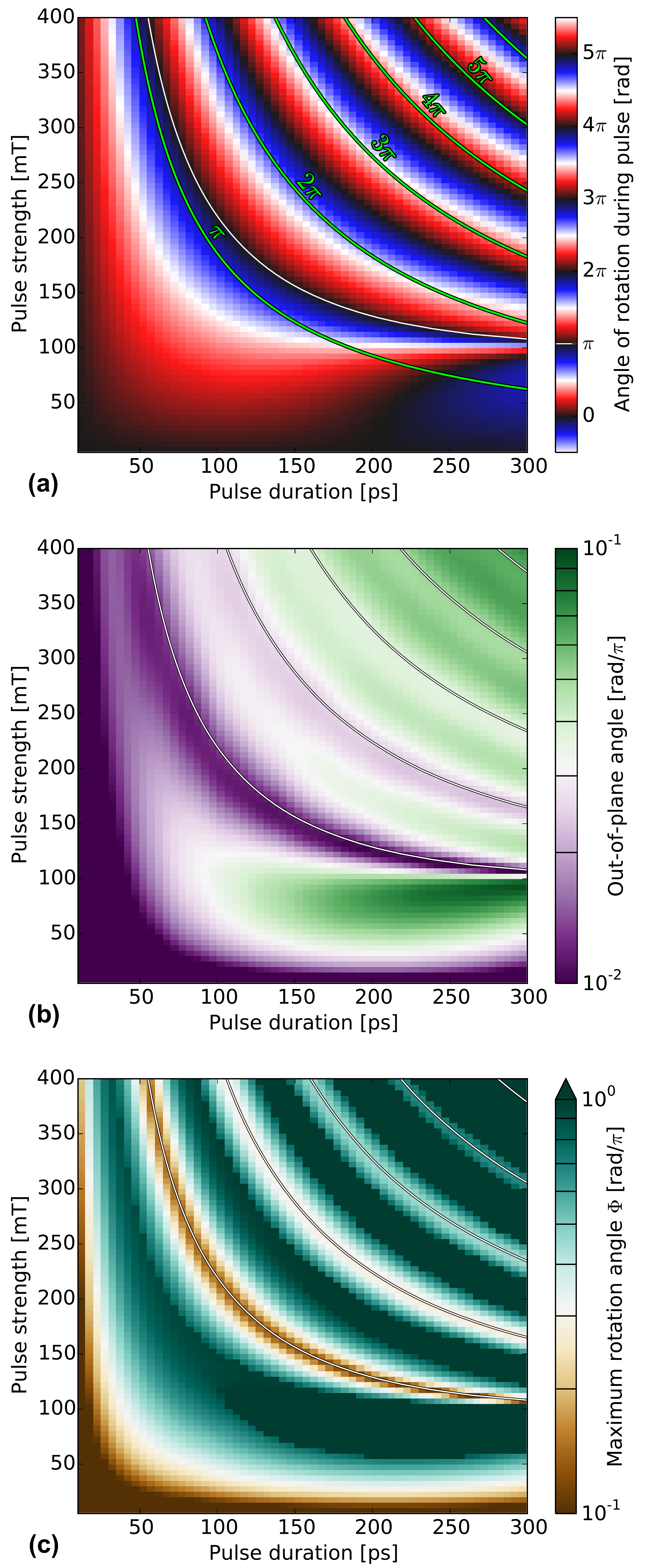}
\caption{(a) Numerically simulated $\Delta\Phi$ of a 33x30 nm SAF particle with layer thicknesses of 5 nm. The green lines labeled with $n\pi$ are the theoretically predicted values \cite{Theory}, for which $\Delta\Phi=n\pi$. (b) Angle of the magnetization with respect to the film-plane, showing a clear minimum around the numerically obtained $\Delta\Phi=n\pi$ (white lines). (c) Maximum magnitude of the in-plane magnetization relaxation, following the pulse-excitation: $\max(\Phi(t\geq0)-\Phi(t=\infty))$. The data show a very significant reduction of the ringing effect when the numerical criterion $\Delta\Phi=n\pi$ is satisfied (contour-lines).
\label{fig:fig2}}
\end{figure}

When analyzing the angle of rotation of the magnetization during the excitation as a function of the pulse strength and duration we find that the analytical prediction is quite accurate for short strong pulses ($\Delta t < 150$ ps and $ H_0 > 150$ mT), as shown in figure \ref{fig:fig2}a for a 33x30 nm trilayer. The functional form of the numerically obtained criterion for the $n\pi$ rotation (multiple switching) during the field-pulse agrees well with the theoretically predicted precession \cite{Theory} shown with green lines in fig. \ref{fig:fig2}a. For the theory to hold, the out-of-plane angle of the magnetization ($\theta_{1,2}$) should be small at the end of the pulse, which largely is the case whenever $\Delta\Phi=n\pi$ is satisfied (fig. \ref{fig:fig2}b). The theory also predicts that the amount of post-excitation relaxation is directly related to the angle of the magnetization at the end of the pulse, both in-plane as well as out-of-plane. The simulated magnetization dynamics shown in figure \ref{fig:fig2}c confirm the analytical prediction that the relaxation is minimal if the power/timing is such that the magnetization at the end of the pulse is near the easy axis. Figure \ref{fig:fig2}c also illustrates the limits of applicability of the theory, which strongly dependent on how well the conditions of equations (\ref{eq:pulse_crit}) are satisfied.

We have compared our results for a SAF with the response of a single ferromagnetic particle with the same physical parameters and found that the dipolar coupling of the two layers in the SAF effectively reduces the field strength necessary for dynamic switching by $\sim$15\%. Theoretically, this effect directly follows from putting the interlayer coupling terms ($\gamma$-terms) to zero, in which case significantly stronger and longer field pulses are necessary to satisfy the criteria of eqs. \ref{eq:pulse_crit} to the same degree.

\begin{figure}
\includegraphics[width=0.9\linewidth]{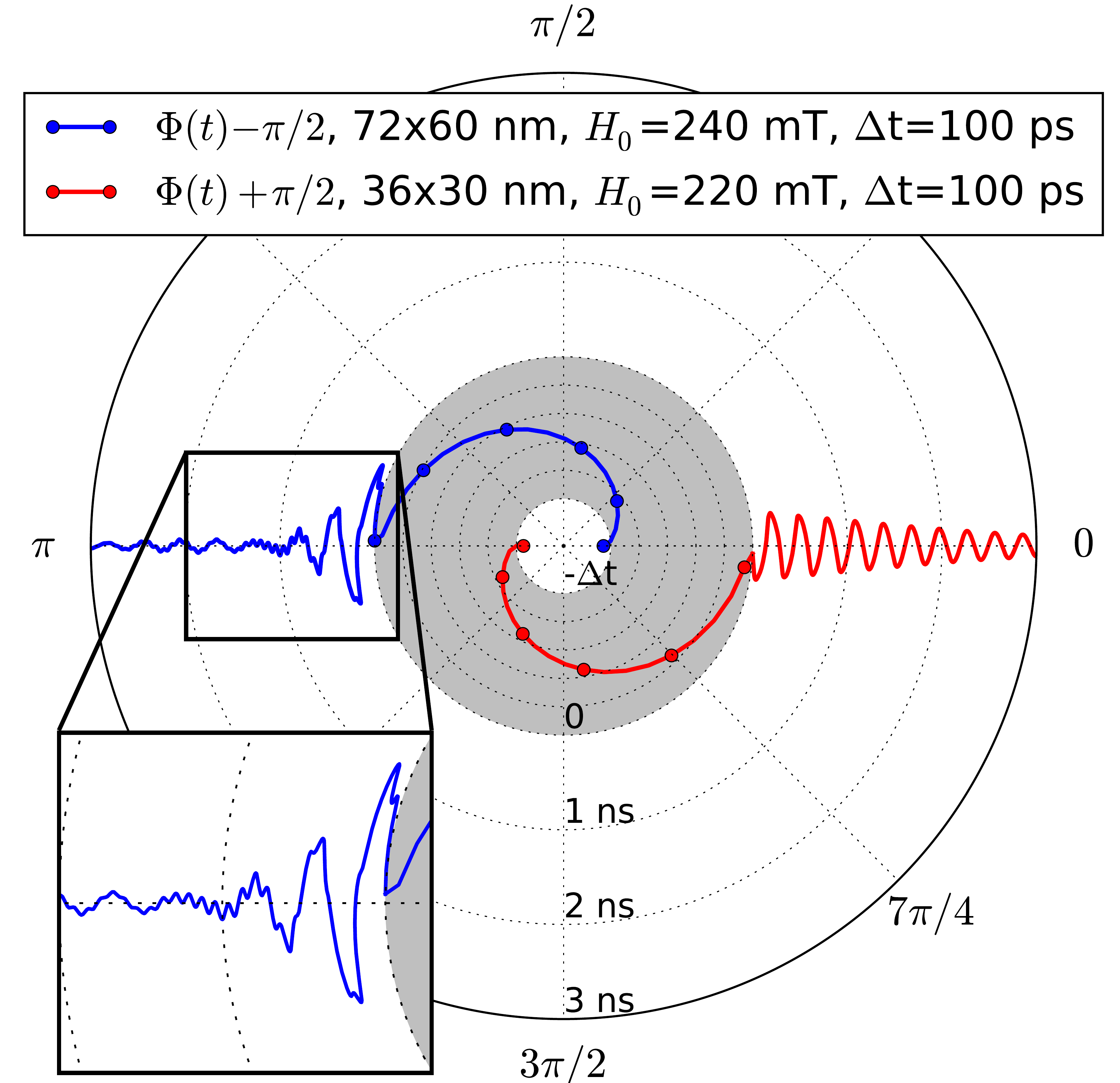}
\caption{\label{fig:fig3} Time evolution of the in-plane magnetization angle $\Phi$ for a 72x60 nm SAF sample, sufficiently large to have spin-waves excited in addition to in-plane rotation (blue), which leads to spin-wave mediated damping. Data for a 36x30 nm SAF sample (red), exhibiting coherent rotation, is shown for comparison. The radial scales are the same as for figure \ref{fig:fig1}b.}
\end{figure}

The results shown so far are for small particles, for which the magnetization is essentially uniform. For larger magnetic particles, such that they can no longer be considered fully single-domain, the spin dynamics can be significantly influenced by spin waves. We have performed simulations for a broad range in lateral sample dimensions (24 nm $\leq\! L_{a} \! \leq 120$ nm) and found that the amount of post-switching relaxation significantly increases with size, even when the criterion of $\Delta\Phi=\pi$ is satisfied. This increase is mainly due to increased spatial nonuniformity as well as the increase of the out-of-plane magnetization at the end of the pulse. Additionally, increasing the aspect-ratio ($L_a/L_b$) effectively increases the energy-barrier for switching to occur. This in turn results in a higher minimal pulse-strength necessary for switching. A larger aspect ratio results in stronger nonuniformity after the pulse.

\begin{figure*}
\includegraphics[width=0.8\linewidth]{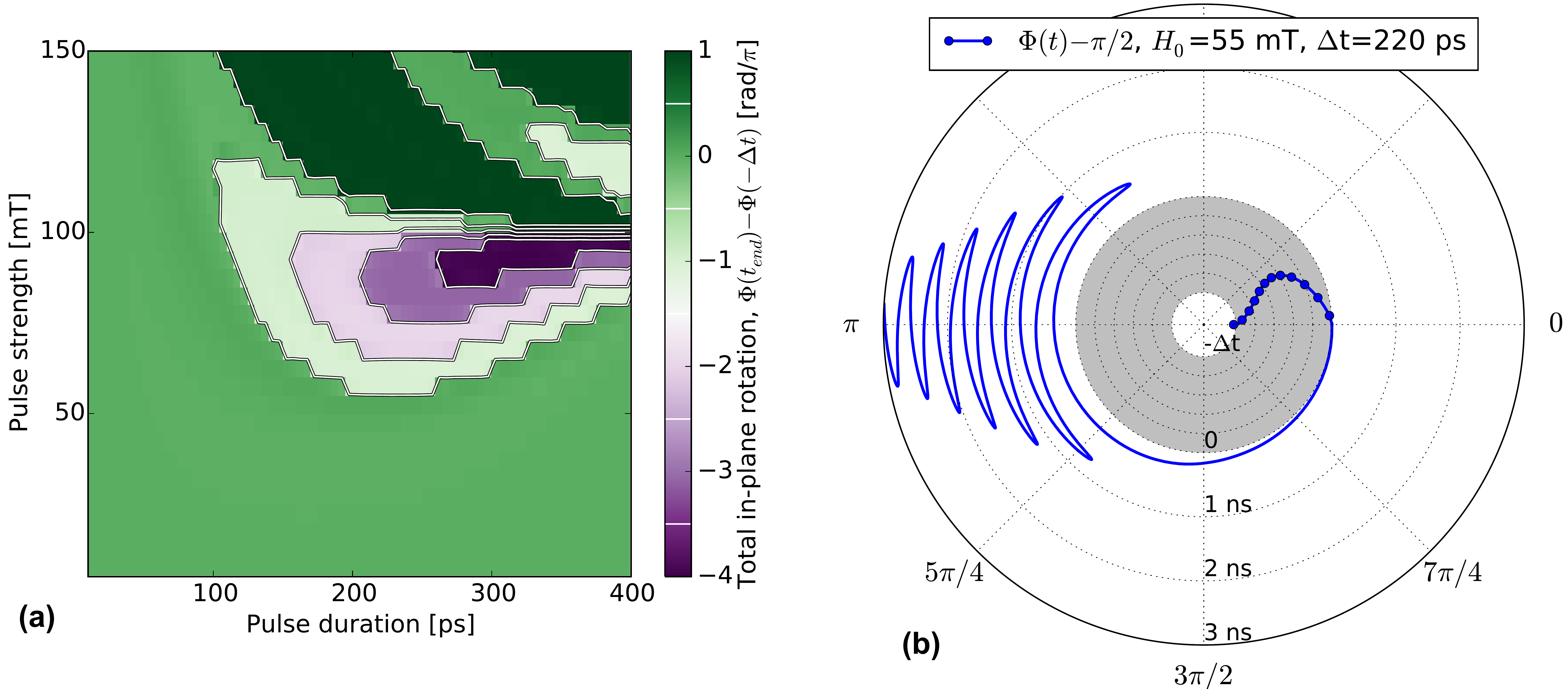}
\caption{Inertial switching for field strengths below the switching-limit for a 33x30 nm SAF sample. If the field pulse is switched off when the direction of the magnetization rotation is opposite to the torque induced by the pulse, the change in inertia is in the direction of the magnetization movement, which can result in inertial switching even for ultra-fast field pulses. (a) Phase-space of the total rotation, from just before the pulse to full relaxation. Clear regions exist of multiple full rotations (multiple sequential switching), taking place well after the field pulse. The number of rotations is directly determined by the angular-velocity of the magnetization at the pulse end. (b) Time-domain evolution of the in-plane angle showing an oscillation during the field pulse and switching inertially, well past the field pulse. The radial scale is the same as in figure \ref{fig:fig1}b.}
\label{fig:fig4}
\end{figure*}

Interestingly, when the pulse duration corresponds to half the period of the spin-wave resonance within the layers, the spin-wave modes are effectively excited. When, additionally, the pulse amplitude is tuned such that $\Delta\Phi=n\pi$, the spin-wave excitation (the so-called magnon-magnon scattering) can dominate the effective magnetic damping in the system, which results in a significantly reduced relaxation time. Figure \ref{fig:fig3} shows such enhanced damping for a 72x60 nm SAF sample, where the resonant-spin-wave excitation greatly speeds up the magnetization relaxation process. For comparison we include the time-domain response for a 36x30 nm sample, with the same pulse-duration and $\Delta\Phi$, exhibiting a long-lived post-switching relaxation. Hence, for optimal spin-wave mediated damping, the magnetization should rotate by $n\pi$ during the action of the excitation pulse having the duration of half the spin-wave period, determined by the specific geometry of the sample. This resonant-spin-wave effect in post-switching magnetization relaxation is effective in samples small enough to be within the single-domain limit quasi-statically, but sufficiently large to exhibit nonuniform spin dynamics - larger than 50 nm in the plane for typical device layouts in our SAF system. 

For weak pulses ($H_0<100$ mT) the applied field torque is insufficient for the magnetization to overcome the shape-anisotropy energy barrier during the pulse.
Our simulations show that for such pulse amplitudes switching can occur inertially via the dynamic energy stored within the SAF particle, enhanced by its acoustical spin resonance \cite{acoust}. 
Figure \ref{fig:fig4}a shows the magnetization angle at the end of the simulation, timed to be well after the excitation with the particle in the final ground state, ($\Phi(t_{end}) - \Phi(-\Delta t) = (n+\frac{1}{2})\pi$, $n\in\mathbb{Z}$ ), as a function of the pulse parameters.
The amplitude and duration are limited to 150 mT and 400 ps respectively to focus on inertial switching. The total angle of rotation is strongly dependent on the amount of inertia the magnetization has directly after the field pulse. Whenever the field is insufficient to directly overcome the in-plane anisotropy, the magnetization precesses about its dynamic-equilibrium axis, which has a finite out-of-plane moment, following the Landau-Lifshitz torque:
\begin{equation}
\frac{d\vec{m}}{dt} = -\gamma \left(\vec{m}\times \vec{H}_{int} + \vec{m}\times(H_{0}\vec{e}_z)\right)-\alpha\vec{m}\times\frac{d\vec{m}}{dt}.
\end{equation}
\noindent Here $\gamma$ is the gyromagnetic ratio; $\vec{H}_{int}$ the effective field minus the Zeeman field; $\alpha$ the damping constant. Taking $m_z$ to be small (good approximation for short pulses), the magnetization vector can be written as $\vec{m}_i=(\,\cos(\varphi_i),\,\sin(\varphi_i),\,0\,)$. Switching off $H_0$ results in an increase of the in-plane angular velocity ($\frac{d\varphi_i}{dt}$) of approximately $-\gamma H_0$. As a result, the fall edge of the field can substantially increase the in plane inertia and, especially if tuned to the acoustical SAF-resonance, produce switching well after the pulse excitation.

Figure \ref{fig:fig4}b illustrates inertial switching for a 33x30 nm particle. Although the field torque is too weak to directly overcome the shape-anisotropy energy barrier, the greatly enhanced magnetic inertia caused by a resonant fall-edge of the field pulse results in switching. Although this switching is far from relaxation-free, the big advantage of this resonant-inertial switching method is that the write pulse energy is significantly lower than for direct dynamical switching. Note that this resonance is not the same as in the case of the in-plane field-pulse as demonstrated in \cite{asym1,thermal1,asym2,asym3}, since in that case the field primarily forces the out-of-phase (optical) rotation in the SAF, which is sub-optimal since in that case the desired acoustical switching is secondary, induced via damping rather than direct torque. 

In conclusion, extensive numerical analysis is used to illustrate a new method of relaxation-free magnetization switching in nanoscale synthetic antiferromagnets subject to 100 ps range perpendicular-to-the-plane field pulses. Another potentially very useful effect is inertial switching, which can be very efficient in a SAF particle if the timing of the field pulse is tuned to the acoustical resonance in the system. Finally, the spin-wave resonance mediated damping is shown to enhance the post-switching stability suitably selected device layouts. These results should prove useful for designing future ultra-fast memory cells based on synthetic nanomagnets.

\subsection*{Acknowledgements}
\noindent This research was funded by the Swedish Research Council, project number 2014-4548. 

\noindent Most of the simulations were performed on resources 
provided by the Swedish National Infrastructure for Computing (SNIC) 
at PDC Centre for High Performance Computing (PDC-HPC)
\bibliography{IEEEexample}
\end{document}